\documentclass[aip,pop,reprint,showkeys,floatfix,preprintnumbers]{revtex4-1} 
\usepackage[T1]{fontenc}
\usepackage[english]{babel}
\usepackage{natbib}
\usepackage[pdftex]{graphicx}
\usepackage{amsmath}
\usepackage{txfonts}
\usepackage{verbatim}
\usepackage{comment}
\usepackage{setspace}
\usepackage{color}
\usepackage{subcaption}

\newcommand{\Fref}[1]{Figure~\ref{#1}}

\newcommand{\fref}[1]{Fig.~\ref{#1}}

\newcommand{\Eref}[1]{Equation~(\ref{#1})}

\newcommand{\eref}[1]{Eq.~(\ref{#1})}
\newcommand{\erefs}[2]{Eqs.~(\ref{#1})-(\ref{#2})}

\newcommand{\sref}[1]{Sec.~\ref{#1}}

\newcommand{\curlM}[1]{\nabla\times{\boldsymbol{ #1}}} 

\graphicspath{{Figure/}}

\begin{document}

\preprint{To be submitted to Phys. Plasmas}
\preprint{April 2019}

\title{Analysis of fast turbulent reconnection with self-consistent determination of turbulence timescale}
\author{F.~Widmer}
\affiliation{Aix-Marseille University, CNRS, PIIM UMR 7345, Avenue Escadrille Normandie Niemen, 13013 Marseille, France}
\email{fabien.widmer@univ-amu.fr}
\altaffiliation{Max Planck Institute for Solar System Research, Justus-von-Liebig-Weg 3, 37077 G\"ottingen, Germany}
\author{J.~B\"uchner}
\affiliation{Max Planck Institute for Solar System Research, Justus-von-Liebig-Weg 3, 37077 G\"ottingen, Germany}
\author{N.~Yokoi}\thanks{Visiting Researcher at the Nordic Institute for Theoretical Physics (NORDITA).}
\affiliation{Institute of Industrial Science, University of Tokyo, 4-6-1 Komaba, Meguro, Tokyo 153-8505, Japan}

\begin{abstract}
	We present results of Reynolds-averaged turbulence model simulation on the problem of magnetic reconnection. In the model, in addition to the mean density, momentum, magnetic field, and energy equations, the evolution equations of the turbulent cross-helicity $W$, turbulent energy $K$ and its dissipation rate $\varepsilon$ are simultaneously solved to calculate the rate of magnetic reconnection for a Harris-type current sheet. In contrast to previous works based on algebraic modeling, the turbulence timescale is self-determined by the nonlinear evolutions of $K$ and $\varepsilon$, their ratio being a timescale. We compare the reconnection rate produced by our mean-field model to the resistive non-turbulent MHD rate. To test whether different regimes of reconnection are produced, we vary the initial strength of turbulent energy and study the effect on the amount of magnetic flux reconnected in time.  
\end{abstract}


\maketitle

\section{Introduction \label{intro}}
	The conversion of stored magnetic energy into different forms by magnetic reconnection is a key dynamical process for astrophysics and laboratory plasmas. In the sun, magnetic reconnection is believed to be the triggering process of solar flares as well as the mechanism responsible for heating the solar corona. Based on linear theory, the rate of energy conversion is given by the inverse magnetic Reynolds number $R_m$ (Sweet--Parker scaling) which is large for astrophysical plasmas. It turns out that the related reconnection rate is very small and can not explain the speed at which dynamical events in the solar corona take place. Several processes for collisionless plasmas have been proposed to increase the energy conversion rate.\cite{Schindler,2007PPCF...49..325B,2005PhPl...12f2902S,2005SSRv..121..237B,terasawa1983hall,JGRA:JGRA15381} For MHD, the plasmoid instability can increase the rate of magnetic reconnection.\cite{Loureiro:2007gv,PhysRevLett.105.235002,bhattacharjee2009fast,2015arXiv151201520H} However, turbulence is ubiquitous in astrophysical plasmas and may play a very important role for reaching fast magnetic reconnection.\cite{Matthaeus_Lamkin_1985,lazarian1999reconnection,kowal2009numerical} \citet{lazarian1999reconnection} proposed that the effective thickness of the diffusion layer is largely increased by the stochastic movement of the magnetic field lines. The argument has been shown not to hold with 3-dimensional simulations of stratified turbulence.\cite{2016arXiv160108167J} A possible reason is that turbulence should be determined simultaneously with the evolution of mean-fields and not be externally imposed. In such situations, the statistical description of turbulence in mean-field theory is adequate/appropriate.

	The range of astrophysical parameters for realistic numerical simulations is not reachable in a foreseeable future and turbulence has to be modeled. Even though several approaches to turbulence model exist,\cite{1980opp..bookR....K,2001MeScT..12.1745S} we concentrate our work on a mean-field turbulence model. The model chosen is a Reynolds-averaged turbulence model where turbulence is self-generated and -sustained by inhomogeneities of the mean-fields.\cite{Yokoi1} Through the model, mean-field MHD equations are solved together with transport equations modeling subgrid turbulent motions. Statistical quantities such as the turbulent cross-helicity (cross-correlation between magnetic-field and velocity-field fluctuations), the turbulent energy and its dissipation rate are the building blocks of the model. The dynamical balance between the turbulent energy and cross-helicity has been presented to produce fast reconnection.\cite{Yokoi1,YokCISM} Such mean-field model has already been confirmed in the context of magnetic reconnection through filtering procedure of high resolution simulations of plasmoid instability in MHD.\cite{Widmer2} Other confirmations such as the expression of the electromotive force by direct numerical simulations (DNS)\cite{YokBal} or the reproducibility of the Alfv\'enicity of the solar wind,\cite{2007PhPl...14k2904Y} strengthens the validity of the present mean-field approach of turbulence.

	An important point in turbulence modeling is the determination of the turbulence timescale which is related to the closure scheme of the turbulence model under consideration. The original turbulence model for magnetic reconnection had been proposed in order that the timescale of turbulence may be self-consistently determined through the turbulence dynamics.\cite{Yokoi1} However, as a first step, a simplified algebraic model for the turbulence timescale was numerically solved.\cite{Yokoi3} There different turbulence timescales used as an external parameter led to different results. Depending on the initial value of the turbulence timescale, two regimes of energy conversion have been found to be \textit{laminar} and \textit{fast turbulent} while a third one was determined to be a \textit{turbulent diffusion} not favorable to a situation of fast reconnection.\cite{Yokoi3} These results have been confirmed and extended to force-free and Harris-type current sheets with out-of-plan guide magnetic field.\cite{Widmer1} However, as noted in the previous work,\cite{Yokoi1,Yokoi3} the timescale of turbulence should be determined directly by the nonlinear dynamics of turbulence. Its spatiotemporal evolution should be determined either by the transport equation of timescale or by some quantities representing it. For instance, in the Kolmogorov picture of hydrodynamic turbulence, the ratio of the turbulent energy with its dissipation rate corresponds to a timescale. Solving simultaneously transport equations for the turbulent energy and its dissipation rate by numerical methods gives a spatiotemporal distribution of the turbulence timescale. In this paper, we solve such transport equations and present results confirming that fast reconnection is obtained by the model. On the other hand, the turbulent diffusive regime is not recovered even for large initial amplitude of turbulence.

\section{Modeling turbulence transport and timescale}
\subsection{Mean-Field Turbulence Model}
In the Reynolds-averaged turbulence model, any instantaneous physical quantity $f$ is divided into a mean $F=\left<f\right>$ and a fluctuation around it, $f'$, as
\begin{equation}
	f=F+f',\;\; F=\left<f\right>
\end{equation}
with 
\begin{subequations}
\begin{eqnarray}
	f &=& (\boldsymbol{v}, \mbox{\boldmath$\omega$}, \boldsymbol{b}, \boldsymbol{j}, \rho, p, h),\\
	F &=& (\boldsymbol{V}, \mbox{\boldmath$\Omega$}, \boldsymbol{B}, \boldsymbol{J}, \overline{\rho}, P, \overline{h}),\\
	f' &=& (\boldsymbol{v}', \mbox{\boldmath$\omega$}', \boldsymbol{b}', \boldsymbol{j}', \rho', p', h'),
\end{eqnarray}
\end{subequations}
where $\boldsymbol{v}$ is the velocity, $\mbox{\boldmath$\omega$} (= \nabla \times \boldsymbol{v})$ the vorticity, $\boldsymbol{b}$ the magnetic field, $\boldsymbol{j} (= \nabla \times \boldsymbol{b})$ the electric-current density, $\rho$ the density, $p$ the pressure, $h$ the internal energy. Here, $\left< {\cdots} \right>$ is the ensemble average satisfying the Reynolds' rules: $\langle {\langle {f} \rangle }\rangle = \langle {f} \rangle$, $\langle {f'} \rangle = 0$, $\langle {f \langle {f} \rangle} \rangle = \langle {f} \rangle \langle {f} \rangle$.\cite{2015LRCA....1....2S,2001MeScT..12.1745S} 
Under the ensemble averaging, the mean-field induction equation reads
\begin{equation}
	\frac{\partial \boldsymbol{B}}{\partial t} 
	= \nabla\times(\boldsymbol{V} \times \boldsymbol{B} +\boldsymbol{\cal{E}})
	+ \eta \nabla^2 \boldsymbol{B},
	\label{eq:Indu_Mean}
\end{equation}
where $\boldsymbol{\cal{E}}$ is the turbulent electromotive force (EMF) defined by
\begin{equation}
	\boldsymbol{\cal{E}}
	\equiv\left<\boldsymbol{v}'\times \boldsymbol{b}'\right>.
	\label{eq:EMF0}
\end{equation}
The EMF solely represents the direct effects of turbulence on the mean induction equation. 
From the equations of the fluctuation fields, the evolution equation for $\boldsymbol{\cal{E}}$ is written as
\begin{subequations}\label{eq:EMFTensor}
\begin{align}
	\frac{D{\cal{E}}_i}{Dt}
	= \, &\left<b'_k\epsilon_{ijk}\partial_\ell b'_j /\overline{\rho}
	-v'_k \epsilon_{ijk}\partial_\ell v'_j \right> B_\ell \label{eq:EMFH00}\\
	&-\left<v'_\ell v'_k + b'_\ell b'_k/(\mu \overline{\rho}) \right> \mu \epsilon_{ijk}\partial_\ell B_j \label{eq:EMFK00}\\
	&+\left<v'_\ell b'_k + b'_\ell v'_k\right>\epsilon_{ijk} \partial_\ell V_j+{\rm{H.T.}}, \label{eq:EMFW00}
\end{align}
\end{subequations}
where $D/DT (\equiv \partial /\partial t + \boldsymbol{V} \cdot \nabla)$ is the Lagrange or material derivative, and $\rm{H.T.}$ stands for higher order terms whose detailed expressions are suppressed here.\cite{Yokoi4,Yok_Dens2} 

If we approximate, for the sake of simplicity, that the inhomogeneity of the fluctuations $f' = (\boldsymbol{v}', \boldsymbol{b}')$ along the mean magnetic field can be represented by the curls of $f'$ as
\begin{subequations}
\begin{equation}
	\left\langle {\epsilon_{ijk} B_\ell (\partial_\ell f'_j) f'_k} \right\rangle
	= \frac{1}{3} \delta_{\ell i} B_\ell  \left\langle {\epsilon_{mjk} (\partial_m f'_j) f'_k} \right\rangle,
\end{equation}
and turbulence is isotropic as
\begin{equation}
	\left\langle {f'_i g'_j} \right\rangle
	= \frac{1}{3} \delta_{ij} \left\langle {f'_\ell g'_\ell} \right\rangle
\end{equation}
\end{subequations}
with $g = (\boldsymbol{v}', \boldsymbol{b}')$,
\begin{subequations}
	\begin{align}
	\frac{D{\cal{E}}_i}{Dt}
	= \, &\frac{1}{3}\left<b'_k \epsilon_{k\ell j}\partial_\ell b'_j /\overline{\rho}
	-v'_k \epsilon_{k\ell j}\partial_\ell v'_j \right> B_i \label{eq:EMFH0}\\
	&-\frac{1}{3}\left<v'_k v'_k + b'_k b'_k /(\mu \overline{\rho}) \right> 
		\mu \epsilon_{i\ell j}\partial_\ell B_j 
	\label{eq:EMFK0}\\
	&+\frac{2}{3}\left<v'_k b'_k \right> \epsilon_{i\ell j}\partial_\ell V_j+{\rm{H.T.}}
	\label{eq:EMFW0}
\end{align}
\end{subequations}
This suggests that the turbulent electromotive force \mbox{\boldmath$\cal{E}$} is estimated as
\begin{equation}
	\langle{{\boldsymbol{v}}'\times{\boldsymbol{b}'}}\rangle 
	= \alpha \boldsymbol{B}
	- \beta \mu \boldsymbol{J} 
	+\gamma {\bf{\Omega}}.
	\label{eq:Emf}
\end{equation}
Here, the transport coefficients $\alpha$, $\beta$, and $\gamma$ are determined by the statistical properties of turbulence. Equation~(7) suggests that they are expressed as
\begin{subequations}\label{eq:alpha_beta_gamma_simplest_model}
\begin{eqnarray}
	\alpha &=& \tau_\alpha H,\\
	\beta &=& \tau_\beta K,\\
	\gamma &=& \tau_\gamma W,
\end{eqnarray}
\end{subequations}
where $\tau_s$ with $s = (\alpha, \beta, \gamma)$ is the timescales of turbulence, and the turbulent magnetohydrodynamic (MHD) energy $K$, the turbulent residual helicity $H$, and the turbulent cross helicity $W$ are defined by
\begin{subequations}\label{eq:turb_stat_quantities}
\begin{eqnarray}
H&=& \left<-\boldsymbol{v}'\cdot\boldsymbol{\omega}' + \boldsymbol{b}'\cdot\boldsymbol{j}'/\overline{\rho} \right>,\\
K&=&\left<\boldsymbol{v}'^2 + {\boldsymbol{b}'^2}/{(\mu \overline{\rho})}\right>/2,\label{Kmath}\\
W&=&\left<{\boldsymbol{v}'\cdot\boldsymbol{b}'}\right>.\label{Wmath}
\end{eqnarray}
\end{subequations}

	As we have just seen above, even with the simplest manipulation of the fluctuation equations we get a basic expression of the EMF [Eq.~(\ref{eq:Emf})] with its transport coefficients $\alpha$, $\beta$, and $\gamma$ [Eq.~(\ref{eq:alpha_beta_gamma_simplest_model})]. In order to obtain more suitable expressions, we have to adopt more elaborated analysis of the MHD turbulence. With the aid of a closure theory of inhomogeneous MHD turbulence,\cite{1990PhFlB...2.1589Y,Yokoi4,Yok_Dens1} the theoretical analytical expression of $\boldsymbol{\cal{E}}$ [Eq.~(\ref{eq:EMF0})] was investigated. On the basis of the theoretical results, the model expression for the EMF was proposed.
In the model, the $\alpha$, $\beta$ and $\gamma$ coefficients are expressed in terms of Green functions which depend themselves on time.\cite{1990PhFlB...2.1589Y,Yokoi4,Yok_Dens1}
A set of elaborated expressions for the transport coefficients of turbulence takes the form:
\begin{subequations}\label{eq:CoeffGreen}
\begin{eqnarray}
	\lefteqn{
	\alpha = \int\,\mbox{d}\boldsymbol{k}\int\limits_{-\infty}^{t}\mbox{d}\tau' G(k,\boldsymbol{x};\tau,\tau',t) }\nonumber\\
	&\hspace{20pt}\times[-H_{vv}(k,\boldsymbol{x};\tau,\tau',t)+H_{bb}(k,\boldsymbol{x};\tau,\tau',t]),
	\label{eq:AlphaGreen}
\end{eqnarray}
\begin{eqnarray}
	\lefteqn{
	\beta = \int\,\mbox{d}\boldsymbol{k}\int\limits_{-\infty}^{t}\mbox{d}\tau' G(k,\boldsymbol{x};\tau,\tau',t) }\nonumber\\
	&\hspace{20pt}\times[Q_{vv}(k,\boldsymbol{x};\tau,\tau',t)+Q_{bb}(k,\boldsymbol{x};\tau,\tau',t]),\label{eq:BetaGreen}
\end{eqnarray}
\begin{eqnarray}
	\lefteqn{
	\gamma = \int\,\mbox{d}\boldsymbol{k}\int\limits_{-\infty}^{t}\mbox{d}\tau' G(k,\boldsymbol{x};\tau,\tau',t) 
	}\nonumber\\
	&\hspace{20pt} \times[Q_{vb}(k,\boldsymbol{x};\tau,\tau',t)+Q_{bv}(k,\boldsymbol{x};\tau,\tau',t]),\label{eq:GammaGreen}
\end{eqnarray}
\end{subequations}
where $G$ is the Green's function of turbulence and $Q_{vv}$, $Q_{bb}$, $H_{vv}$, $H_{bb}$, $Q_{vb}$ and $Q_{bv}$ are the spectral functions in the wavenumber space. The derivation and physical meaning of Eq.~(\ref{eq:CoeffGreen}) can be found in\citep{Yokoi4} As we will see in the following subsection, Eq.~(\ref{eq:CoeffGreen}) is a natural generalization of the simplest expressions given by Eq.~(\ref{eq:alpha_beta_gamma_simplest_model}).

In view of Eqs.~(\ref{eq:alpha_beta_gamma_simplest_model}) and (\ref{eq:CoeffGreen}), it is obvious that we have to adopt appropriate timescales of turbulence in order to properly model the turbulent transport coefficients $\alpha$, $\beta$, and $\gamma$.

\subsection{Turbulence transport and timescale}
Depending on the level of the turbulence closure scheme, there are several ways to express the effective transport coefficients due to turbulence. Let us see this point, for example, in the case of  the eddy or turbulent viscosity. The following arguments are based on a rather simple picture of homogeneous isotropic turbulence, but will provide a perspective on the relationship of the timescale of turbulence and turbulent transport. 

	The simplest way of describing the turbulent momentum transport may be just to treat the turbulent viscosity as a parameter. In this level of description, the eddy viscosity is given as
\begin{equation}
	\nu_{\rm{T}} = \nu_{\rm{T}0},
	\label{eq:nuT_parameter}
\end{equation}
with $\nu_{{\rm{T}0}}$ being a parameter, and the value of $\nu_{{\rm{T}}0}$ is determined in a heuristic manner. 

	The next higher level of description might be given by the so-called mixing-length modeling, where the transport coefficient is expressed by the characteristic velocity and length scales of turbulence, $u$ and $\ell$, as
\begin{equation}
	\nu_{\rm{T}} \sim u \ell \sim u^2 \tau,
	\label{eq:nuT_mixing}
\end{equation}
where $\tau$ is the characteristic timescale of turbulence and $\ell$ is called the mixing length, which is chosen in a heuristic manner based on the physical arguments. In this sense, the choice of $\ell$ is not directly based on the turbulence dynamics.

	In the present work, we use the results by a more elaborated closure theory of inhomogeneous turbulence, the two-scale direct-interaction approximation (TSDIA). In this framework, the transport coefficients are expressed in terms of the Green's and spectral functions. In this formulation, the eddy viscosity $\nu_{\rm{T}}$ is expressed as
\begin{equation}
	\nu_{\rm{T}}
	= \frac{7}{15} \int \!d{\bf{k}} \int_{-\infty}^{t} \!\!\!d\tau_1
	G({\bf{k}}, {\bf{x}};\tau,\tau_1) Q({\bf{k}}, {\bf{x}};\tau,\tau_1,t),
	\label{eq:nuT_by_tsdia}
\end{equation}
where $G({\bf{k}}, {\bf{x}}; \tau, \tau_1)$, and $Q({\bf{k}}, {\bf{x}}; \tau, \tau_1, t)$ are the Green's and energy spectral functions, respectively. Since the Green's function represents the weight how much the past states affect the present one, it basically gives information of timescale of turbulence. In a simpler case where the spectral function $Q$ is not affected by the past time, the time integral of the Green's function is separately calculated. Then Eq.~(\ref{eq:nuT_by_tsdia}) is reduced to the spectral expression of the eddy viscosity
\begin{equation}
	\nu_{\rm{T}} = \int_{k_0}^{\infty} \!\!\!dk\; \tau(k) E(k),
	\label{eq:nuT_spect_exp}
\end{equation}
where $k_0=(2\pi/\ell)$ is the wavenumber of the energy-containing eddy or mixing length $\ell$, $\tau$ is the timescale of turbulence, and $E(k)$ is the energy spectrum, showing how much energy is contained in the scale represented by wavenumber $k$. Note that if $\tau$ is represented by motions of the largest or energy-containing scales, and does not depend on scale, Eq.~(\ref{eq:nuT_spect_exp}) reads to the simple mixing-length expression for the eddy viscosity [Eq.~(\ref{eq:nuT_mixing})]. From these arguments, we see that Eq.~(\ref{eq:nuT_by_tsdia}) is a much more elaborated expression capturing more complicated spatiotemporal properties of turbulence.

As we observed in Eqs.~(\ref{eq:nuT_parameter})-(\ref{eq:nuT_by_tsdia}), whatever levels of description might be, the timescale of turbulence is directly related to the turbulent transport properties, and is one of the most important ingredients of turbulence modeling. There are several timescales in turbulence dynamics; the eddy distortion or turnover time, the spectral transfer time, the timescale associated with the decay of the triple correlations, etc. In the Kolomogorov's picture for the homogeneous hydrodynamic turbulence, all those timescales are the same, and we substantially have only one timescale.

	We assume the Kolmogorov's scaling for the energy spectrum:
\begin{equation}
	E(k) = C_{\rm{K}} \varepsilon^{2/3} k^{-5/3},
	\label{eq:kolmo_scaling}
\end{equation}
where $C_{\rm{K}}$ is the Kolmogorov constant of the magnitude of $O(1)$ and $\varepsilon$ is the energy dissipation rate, which is equivalent to the energy flux or transfer rate in the wave-number space in the Kolmogorov picture of turbulence.

	Substituting Eq.~(\ref{eq:kolmo_scaling}) into Eq.~(\ref{eq:nuT_spect_exp}), we estimate the turbulent energy $K$ as
\begin{equation}
	K
	= \int_{k_0}^{\infty}\!\!\!E(k)\; dk
	= \int_{k_0}^{\infty} \!\!\! C_{\rm{K}} \varepsilon^{2/3} k^{-5/3}\; dk
	\sim \varepsilon^{2/3} k_0^{-2/3}.
	\label{eq:K_kolmo_scaling}
\end{equation}
With this estimate, the eddy turnover time $\tau_{\rm{E}}$ is expressed in terms of the turbulent energy ($K \sim u^2$) and its dissipation rate as
\begin{equation}
	\tau_{\rm{E}} \sim \ell / u \sim K / \varepsilon.
	\label{eq:tau_E}
\end{equation}
Note that Eq.~(\ref{eq:tau_E}) is also obtained directly from the Kolmogorov's four-fifth law $u \sim \delta u \sim (4/5)(\varepsilon \ell)^{1/3}$. In the Kolomogorov's picture, the energy dissipation rate $\varepsilon$ is equivalent to the energy transfer rate: how much the turbulent energy fluxes from larger to smaller scales.

In the presence of the magnetic field, in addition to the eddy turnover time $\tau_{\rm{E}}$, the timescale associated with the Alfv\'{e}n-wave propagation along the magnetic field may play an important role in turbulence evolution. This timescale, the Alfv\'{e}n time $\tau_{\rm{A}}$, is defined as
\begin{equation}
	\tau_{\rm{A}} \sim (k V_{\rm{A}})^{-1},
	\label{eq:alfven_time}
\end{equation}
where $V_{\rm{A}} [=|\boldsymbol{b}| / (\mu \rho)^{1/2}]$ is the Alfv\'{e}n speed of the magnetic field $\boldsymbol{b}$, and $k$ is the wavenumber ($\mu$: magnetic permeability, $\rho$: density). \\

In the Alfv\'{e}n-wave turbulence, the energy transfer is caused solely by the interaction of two Alfv\'{e}n-wave packets propagating in opposite directions along the magnetic field. Each interaction time is the Alfv\'{e}n time $\tau_{\rm{A}}$, which is typically short compared with the eddy turnover time ($\tau_{\rm{A}} \ll \tau_{\rm{E}}$). In order to produce the same amount of energy transfer as in the hydrodynamic turbulence, where the eddy turnover time is $\tau_{\rm{E}}$, many ($\tau_{\rm{E}} / \tau_{\rm{A}}$ times) Alfv\'{e}n-wave interaction events are needed. This suggests that the energy transfer time for the scale represented by the wavenumber $k$, $\tau_{\rm{AE}}$, in the Alfv\'{e}n wave turbulence is longer than the hydrodynamic counterpart $\tau_{\rm{E}}$ as
\begin{equation}
	\tau_{\rm{AE}}
	\sim (\tau_{\rm{E}} / \tau_{\rm{A}}) \tau_{\rm{E}}.
	\label{eq:alfven_en_tr_time}
\end{equation}\\

We assume the spectrum for the inertial range in Alfv\'{e}n-wave turbulence is the Iroshnikov--Kraichnan type:
\begin{equation}
	E(k) = C_{\rm{IK}} (\varepsilon V_{\rm{A}})^{1/2} k^{-3/2},
	\label{eq:IK_scaling}
\end{equation}
where $C_{\rm{IK}}$ is the Iroshnikov--Kraichnan constant of the magnitude of $O(1)$. With this spectrum, the turbulent energy in Alfv\'{e}nic turbulence is estimated as
\begin{equation}
	K = \int_{k_0}^\infty \!\!\!E(k)\; dk
	= \int_{k_0}^\infty \!\!\! C_{\rm{IK}} (\varepsilon V_{\rm{A}})^{1/2} k^{-3/2} dk
	\sim (\varepsilon V_{\rm{A}})^{1/2} k_0^{-1/2},
	\label{eq:K_IK}
\end{equation}
where $k_0$ is the characteristic wavenumber in the energy containing region or largest scale of turbulent motion. With this estimate, the Alfv\'{e}n time can be connected to the eddy turnover time in terms of the turbulent energy and the mean magnetic-field energy as
\begin{equation}
	\tau_{\rm{A}}
	\simeq (k_0 V_{\rm{A}})^{-1}
	\simeq \left( {
		\frac{\varepsilon V_{\rm{A}}}{K^2} V_{\rm{A}}
	} \right)^{-1}
	= \frac{K}{V_{\rm{A}}^2} \frac{K}{\varepsilon}
	= \frac{K}{V_{\rm{A}}^2} \tau_{\rm{E}}.
	\label{eq:tauA_tauE_rel}
\end{equation}
This should be compared with Eq.~(\ref{eq:alfven_en_tr_time}). Equation~(\ref{eq:tauA_tauE_rel}) suggests that the relative magnitude of the Alfv\'{e}n time to the eddy turnover time is proportional to the ratio of the turbulent energy to the mean magnetic field
\begin{equation}
	\tau_{\rm{A}} V_{\rm{A}}^2 \sim \tau_{\rm{E}} K.
	\label{eq:tauA_tauE_rel_simple}
\end{equation}
If the turbulent energy $K$ is comparable with the mean magnetic energy represented by $V_{\rm{A}}^2$, the Alfv\'{e}n time is comparable with the eddy turnover time $\tau_{\rm{E}}$:
\begin{equation}
	\tau_{\rm{A}} \simeq \tau_{\rm{E}}\;\;\; \mbox{for}\;\;\; K \simeq V_{\rm{A}}^2.
	\label{eq:tauA_sim_tauE}
\end{equation}

\subsection{Synthesized Timescale}
If the energy transfer is caused by both the processes of eddy turnover and Alfv\'{e}n-wave interactions, we need to construct a synthesized timescale $\tau_{\rm{S}}$ from the eddy turnover time $\tau_{\rm{E}}$ and the Alfv\'{e}n energy transfer time $\tau_{\rm{AE}}$. The representative one may be the timescale based on average:
\begin{equation}
	\frac{1}{\tau_{\rm{S}}^p}
	= \frac{1}{\tau_{\rm{E}}^p}  + \frac{1}{\tau_{\rm{AE}}^p}
	\label{tauS_harm_ave}
\end{equation}
with the power index $p$, for which $p=1$ or $2$ is typically adopted. If we adopt $p=1$, corresponding to the harmonic average, it gives a synthesized timescale\cite{2008JTurb...9...37Y}:
\begin{equation}
	\tau_{\rm{S}}
	= \frac{1}{1 + \tau_{\rm{E}}/\tau_{\rm{AE}}} \tau_{\rm{E}}
	\sim \frac{1}{1 + \tau_{\rm{A}}/\tau_{\rm{E}}} \tau_{\rm{E}}
	\sim \frac{1}{1+ {V_{\rm{A}}}^2/K} \tau_{\rm{E}},
	\label{eq:tauS_tauE_rel}
\end{equation}
where use has been made of the relationship between $\tau_{\rm{A}}$ and $\tau_{\rm{AE}}$ [Eq.~(\ref{eq:alfven_en_tr_time})] in the second estimate, and that between $\tau_{\rm{A}}$ and $\tau_{\rm{E}}$ [Eq.~(\ref{eq:tauA_tauE_rel_simple})] in the third estimate.

Equation~(\ref{eq:tauS_tauE_rel}) shows that in the weak magnetic-field limit ($V_{\rm{A}}^2 \ll K$), the synthesized timescale recovers the usual eddy turnover time as
\begin{equation}
	\tau_{\rm{S}}
	\simeq \left( {
		1 - \frac{V_{\rm{A}}^2}{K}
	} \right) \tau_{\rm{E}}
	\simeq \tau_{\rm{E}}
	\simeq \frac{K}{\varepsilon}\;\;\; \mbox{for}\;\;\; V_{\rm{A}}^2 \ll K.
	\label{eq:tauS_small_mag}
\end{equation}
On the other hand, in the strong magnetic-field limit ($V_{\rm{A}}^2 \gg K$), the synthesized timescale is reduced to the Alfv\'{e}n time as
\begin{equation}
	\tau_{\rm{S}}
	\simeq \frac{K}{V_{\rm{A}}^2} \left ( {
		1- \frac{K}{V_{\rm{A}}^2}
	} \right) \tau_{\rm{E}}
	\simeq\frac{K}{V_{\rm{A}}^2} \tau_{\rm{E}}
	\simeq \tau_{\rm{A}}\;\;\; \mbox{for}\;\;\; V_{\rm{A}}^2 \gg K.
	\label{eq:tauS_large_mag}
\end{equation}

In Eq.~(\ref{eq:tauA_sim_tauE}) we saw that the Alfv\'{e}n time is comparable to the eddy turnover time: $\tau_{\rm{A}} \sim \tau_{\rm{E}}$ if the turbulent energy is comparable to the mean magnetic field energy, $K \sim V_{\rm{A}}^2$. In this case, it follows from Eqs.~(\ref{eq:alfven_en_tr_time}), (\ref{eq:tauA_tauE_rel}) and (\ref{eq:tauA_sim_tauE}) that all the timescales; the synthesized, eddy turnover, Alfv\'{e}n times, and Alfv\'{e}n energy transfer time are comparable:
\begin{equation}
	\tau_{\rm{S}} \sim \tau_{\rm{E}} \sim \tau_{\rm{A}} \sim \tau_{\rm{AE}}.
	\label{eq:tauS_compara_mag}
\end{equation}
This means that the synthesized energy transfer rate time $\tau_{\rm{S}}$ can be expressed by the eddy turnover time $\tau_{\rm{E}} (=K/\varepsilon)$, which is eventually the same as the Alfv\'{e}n time $\tau_{\rm{A}}$.

In the simulation with a turbulence model (\sref{sec:EQNEPSi}), the mean-field equations are simultaneously solved with the turbulent-field equations. By solving the dissipation-rate equation ($\varepsilon$ equation) as well as the turbulent energy equation ($K$ equation), we obtain the spatiotemporal evolution of the timescale of turbulence by Eq.~(\ref{eq:tauS_tauE_rel}) or more simply by Eq.~(\ref{eq:tau_E}) or Eq.~(\ref{eq:tauS_compara_mag}). This procedure provides a possibility of more self-consistent estimate of the turbulent transport than any turbulence model in which the timescale $\tau$ is just given as a parameter. Under the assumption that the eddy turnover and the Alfvén time are about the same and that the time integrals of the Green's function  can be separately computed from the turbulent field propagators, the transport coefficients $\alpha$, $\beta$ and $\gamma$ [\eref{eq:CoeffGreen}] are modeled as\cite{1984PhFl...27.1377Y,Yokoi2,2016ApJ...824...67Y}
\begin{subequations}
\begin{eqnarray}
\beta &=& C_\beta\tau K, \label{ModK} \\
\gamma &=& C_{\gamma_t}\tau W, \label{ModW} \\
\alpha &=& C_\alpha\tau H. \label{ModH}
\end{eqnarray}
\end{subequations}
where $C_\beta$, $C_{\gamma_t}$ and $C_\alpha$ are model constants and $\tau$ the timescale of turbulence evolution.
As a simplified approach, we consider that the timescale of turbulence can be dynamically determined as
\begin{equation}
	\tau = {K}/{\varepsilon} \label{eq:Tau}.
\end{equation}
The derivation and physical origins of the model expression \eref{eq:Emf} as well as \erefs{ModK}{ModH} and the timescale of turbulence can be found in Yokoi (2013).\citep{Yokoi4}

\section{Basic equations and numerical implementation\label{Basic}}
\subsection{Mean-field equations}
The GOEMHD3 code \cite{Skala:2014cwa} is used to solve the dimensionless mean-field MHD equations normalised to a mass density $\rho_0$, magnetic field strength $B_0$ and a typical length scale $L_0$ taken to be the half-width of the Harris-type current sheet. 
\begin{equation}
	\frac{\partial\overline{\rho}}{\partial t}
	=-\nabla\cdot(\overline{\rho}\boldsymbol{V}),
	\label{eq:density} 
\end{equation}
\begin{equation}
	\frac{\partial}{\partial t} \overline{\rho}\boldsymbol{V}
	=-\nabla\cdot\left[ \overline{\rho}\boldsymbol{V}\otimes\boldsymbol{V}
	+\frac{1}{2}(P+\boldsymbol{B}^2)\boldsymbol{I} 
	- \boldsymbol{B} \otimes \boldsymbol{B}\right]
	+{\textstyle{\chi}}\nabla^2 \overline{\rho}\boldsymbol{V},
	\label{eq:momentum}
 \end{equation}
 \begin{equation}
	\frac{\partial\boldsymbol{B}}{\partial t}
	= \nabla\times\left(\boldsymbol{V} \times \boldsymbol{B}
	+\langle{{\boldsymbol{v}}'\times{\boldsymbol{b}}}'\rangle\right)
	+\eta \nabla^2 \boldsymbol{B},
	\label{eq:induction}
\end{equation}
\begin{equation}
	\frac{\partial\overline{h}}{\partial t}
	=-\nabla\cdot(\overline{h} \boldsymbol{V})
	+\frac{\gamma_0-1}{\gamma_0 \overline{h}^{\gamma_0-1}}
		(\eta\boldsymbol{J}^2)
	+\chi \nabla^2 \overline{h},
	\label{eq:entropy}
\end{equation}
where $\overline{\rho}$, $\boldsymbol{V}$, $\boldsymbol{B}$ and $\boldsymbol{J}$ are the mean mass density, fluid velocity, magnetic field and current density, respectively. The mean current density is calculated as $\boldsymbol{J}=\nabla \times \boldsymbol{B}/\mu$. The symbol $\overline{h}$ represents the internal energy of the mean-fields and is related to the thermal pressure by the equation of state $P=2h^\gamma_0$ for adiabatic conditions ($\gamma_0=C_V/C_p=5/3$ is the ratio of specific heats).\cite{Widmer2} The resistivity is given by $\eta=10^{-4}$ and the parameter $\chi$ describes the amount of local smoothness used to avoid numerical instabilities. It is defined as $\chi= \chi_0 + \chi_{\rm{loc}}$ where $\chi_0=10^{-3}$ is uniform and constant while $\chi_{\rm{loc}}\ll\chi_0$ is activated locally if necessary.

	The turbulent electromotive force $\langle{{\boldsymbol{v}}'\times{\boldsymbol{b}}}'\rangle$ as well as $\varepsilon$ represents the feedback of MHD turbulence in the mean induction equation. As a first step, the turbulent stress-tensor $\langle {\boldsymbol{v}' \otimes \boldsymbol{v}' -  \boldsymbol{b}' \otimes \boldsymbol{b}'} \rangle$ is neglected in the momentum equation (\ref{eq:momentum}) and only the turbulent electromotive force relates turbulence to the mean fields.

\subsection{Turbulence equations \label{sec:EQNEPSi}}
In addition to the system of mean-field MHD \erefs{eq:density}{eq:entropy}, the governing equation for the turbulent energy $K$, its dissipation rate $\varepsilon$
and the turbulent cross-helicity $W$ are implemented in the GOEMHD3 code and solved as
\begin{equation}
	\frac{\partial K}{\partial t}
	=-\boldsymbol{V}\cdot \nabla K 
	-\boldsymbol{\cal{E}} \cdot\boldsymbol{J}
	+ \frac{\boldsymbol{B}}{\sqrt{\overline{\rho}}}\cdot\nabla W 
	- \varepsilon,\label{eq:KEpsi}
\end{equation}
\begin{equation}
	\frac{\partial W}{\partial t}
	=-\boldsymbol{V}\cdot \nabla W 
	-\boldsymbol{\cal{E}} \cdot \boldsymbol{\Omega}
	+\frac{\boldsymbol{B}}{\sqrt{\overline{\rho}}}\cdot\nabla K 
	- C_W\frac{\varepsilon W}{K},
	\label{eq:WEpsi}
\end{equation}
\begin{equation}
	\frac{\partial \varepsilon}{\partial t}
	= -\boldsymbol{V}\cdot\nabla\varepsilon
	+\frac{\varepsilon}{K}\left(C_{\varepsilon 1}P_K
	-C_{\varepsilon 2}\varepsilon
	+C_{\varepsilon 3} \boldsymbol{B} \cdot \nabla W\right),
	\label{eq:Epsi}
\end{equation}
where $\boldsymbol{\cal{E}}$ is the turbulent electromotive force [\eref{eq:Emf}] and $C_{\varepsilon n} (n=1-3)$ model constants of order $\mathcal{O}(1)$ and are chosen as $C_{\varepsilon 1}=1.4$, $C_{\varepsilon 2}=1.9$ and $C_{\varepsilon 3}=1.0$.\cite{1990PhFlB...2.1589Y} The symbol $P_K$ is the turbulent energy production mechanism given as
\begin{equation}
	P_K\equiv -\boldsymbol{\cal{E}} \cdot\boldsymbol{J}
	= \frac{K}{\varepsilon}\left(C_\beta K\frac{\boldsymbol{J}^2}{\overline{\rho}} 
	- C_{\gamma}W \frac{\boldsymbol{\Omega} \cdot \boldsymbol{J}}{\sqrt{\overline{\rho}}}\right).
	\label{eq:PKEpsilon}
\end{equation}	
The purpose of solving \eref{eq:Epsi} in addition to Eqs.~(\ref{eq:KEpsi}) and (\ref{eq:WEpsi}) and the mean-field MHD \erefs{eq:density}{eq:entropy} is to determine the timescale of turbulence accordingly to the turbulence dynamics.  In such a way, we can test whether turbulence produces different regimes of reconnection as obtained in previous works using a constant-$\tau$ approximation.\cite{Yokoi3,Widmer1} Since such regimes of reconnection were obtained independently of the initial current sheet configuration used, we only consider a Harris-type current sheet. The turbulent helicity related term $\alpha$ is neglected in the present work because there is no symmetry breakage capable to generate it in a Harris-type current sheet without guide magnetic field.\cite{Widmer2}\\

	The geometry is defined by the unit vectors $\bf{e}_y$ (across the current sheets), $\bf{e}_z$ (along the current sheets) and $\bf{e}_x$ (in-plan direction). The resolution is $4\times2048\times 2048$ for a box size of $0.4L_x/L_0\times 80L_y/L_0\times80L_z/L_0$ where the normalizing length $L_0$ is the current sheet half-width. To use periodic boundary conditions, a pair of Harris-type current sheet are initialized as
\begin{equation}
	\boldsymbol{B} 
	= B_0\left\{ {
		\tanh{\left[(y+d)/L_0\right]}-\tanh{\left[(y-d)/L_0\right]}-1
	} \right\} \boldsymbol{e}_z
\end{equation}	
for an asymptotic value of the magnetic field $B_{0}$=1. Both current sheets are located at $\pm 10 L_y/L_0$. Reconnection is triggered by a divergence free perturbation 
\begin{equation}
	\delta\boldsymbol{B}=\frac{b_p}{b_0}\sum\limits_{i=1}^{10}\sin(2\pi i z/L_z){\textbf{e}}_z.
\end{equation}
with $b_p/b_0=10^{-3}$.
The initial zero flow condition ($\boldsymbol{V}=0$) gives $W_0=0$ as an initial condition for the cross-helicity. Hence, the initial balance for turbulence is
given by
\begin{equation}
	P_{K_0}-\varepsilon_0=0.
	\label{eq:TurbBalance}
\end{equation}
Here $P_{K_0}$ is the initial turbulence energy
production given by
\begin{equation}
	P_{K_0}= \frac{C_\beta K_0^2}{\varepsilon_0}\frac{\boldsymbol{J}^2}{\overline{\rho}}.
	\label{eqPK0}
\end{equation}
\Eref{eq:TurbBalance} provides an initial condition for the turbulent dissipation rate. 
It depends only on the initial value of the turbulence energy $K_0$ which is varied. The initial strength of the turbulence energy $K_0$ and its dissipation rate $\varepsilon_0$ are given as
\begin{equation}
	K_0 =K_{\rm{init}},
\end{equation}
\begin{equation}
	\varepsilon_0 = \sqrt{C_\beta} K_0 \left|\boldsymbol{J}_0\right|,
	\label{eq:Epsi0}
\end{equation}
where $\mu_0\boldsymbol{J}_0=\curlM{B}_0$. 

\begin{figure}
	\begin{subfigure}[t]{0.5\linewidth}
	\centering
		\includegraphics[width=0.9\textwidth,keepaspectratio]{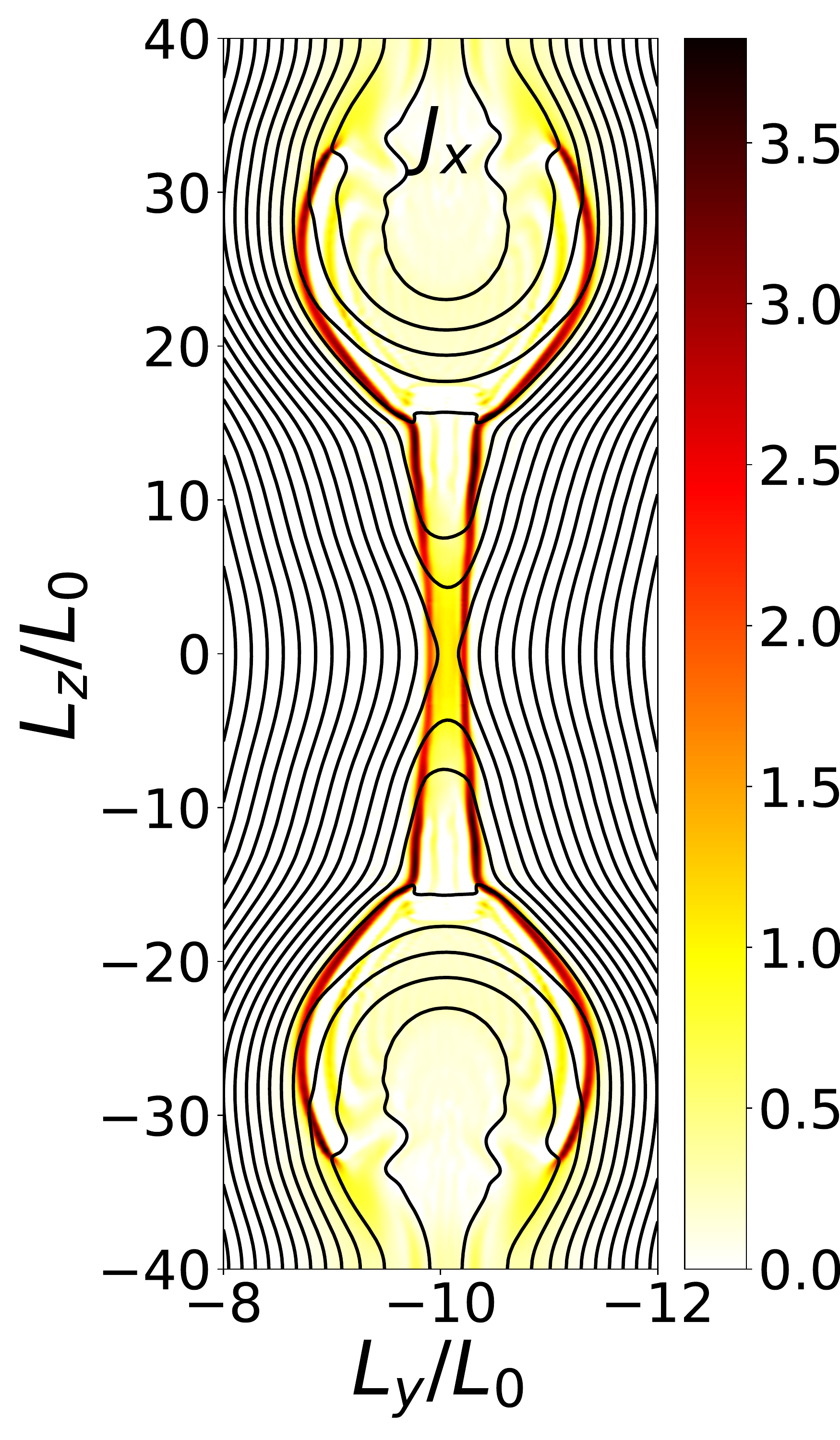} 
	\caption{Mean current density $\boldsymbol{J}$}
	\label{fig:MeanJEp}
       \end{subfigure}~
       \vspace{10pt}
       \begin{subfigure}[t]{0.5\linewidth}
	\centering
	\includegraphics[width=0.875\textwidth,keepaspectratio]{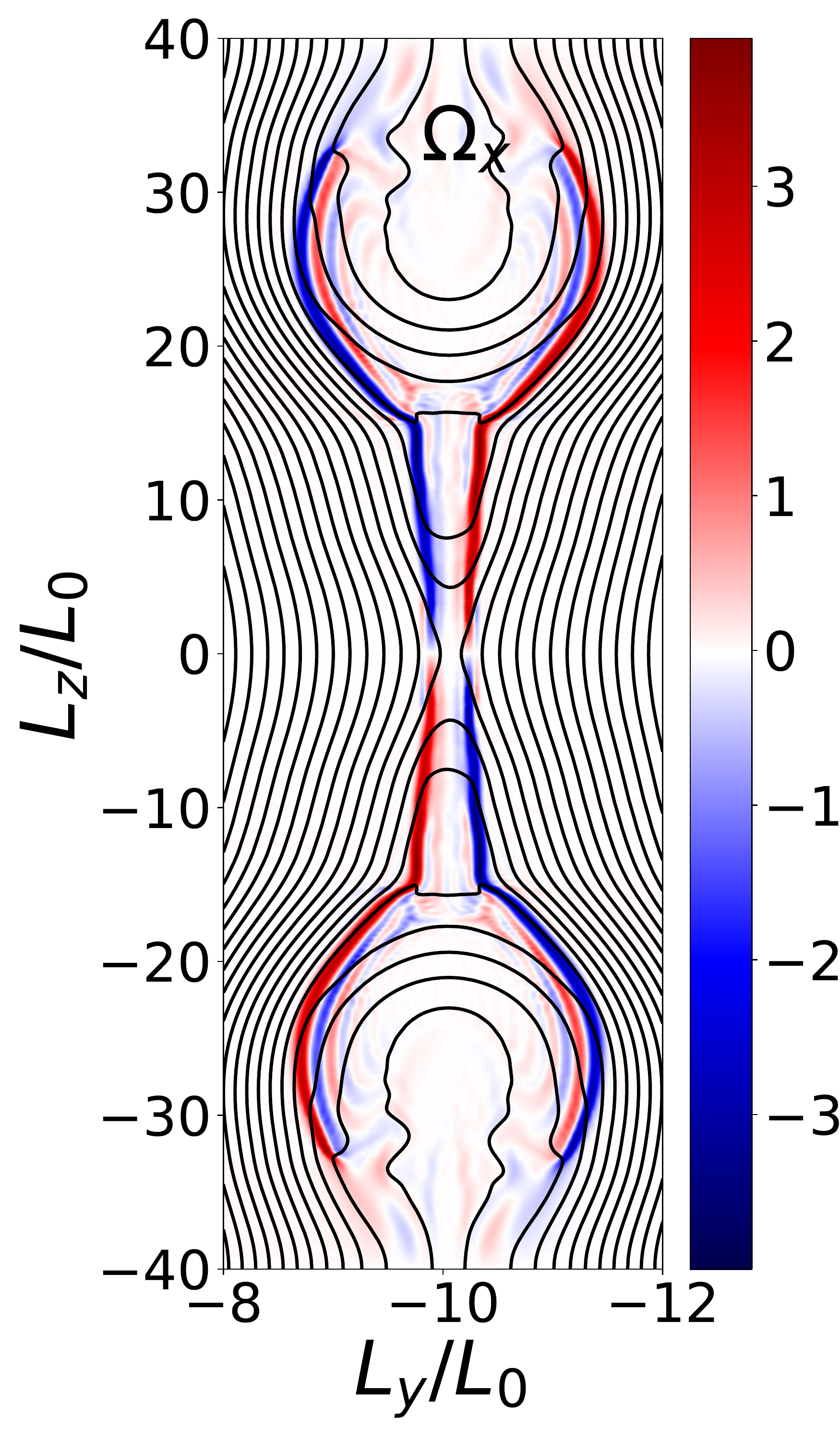}
	\caption{Mean vorticity $\boldsymbol{\Omega}$}
	\label{fig:MeanOEp}
      \end{subfigure}\\
      \begin{subfigure}[t]{0.5\linewidth}
	\centering
	\includegraphics[width=0.9\textwidth,keepaspectratio]{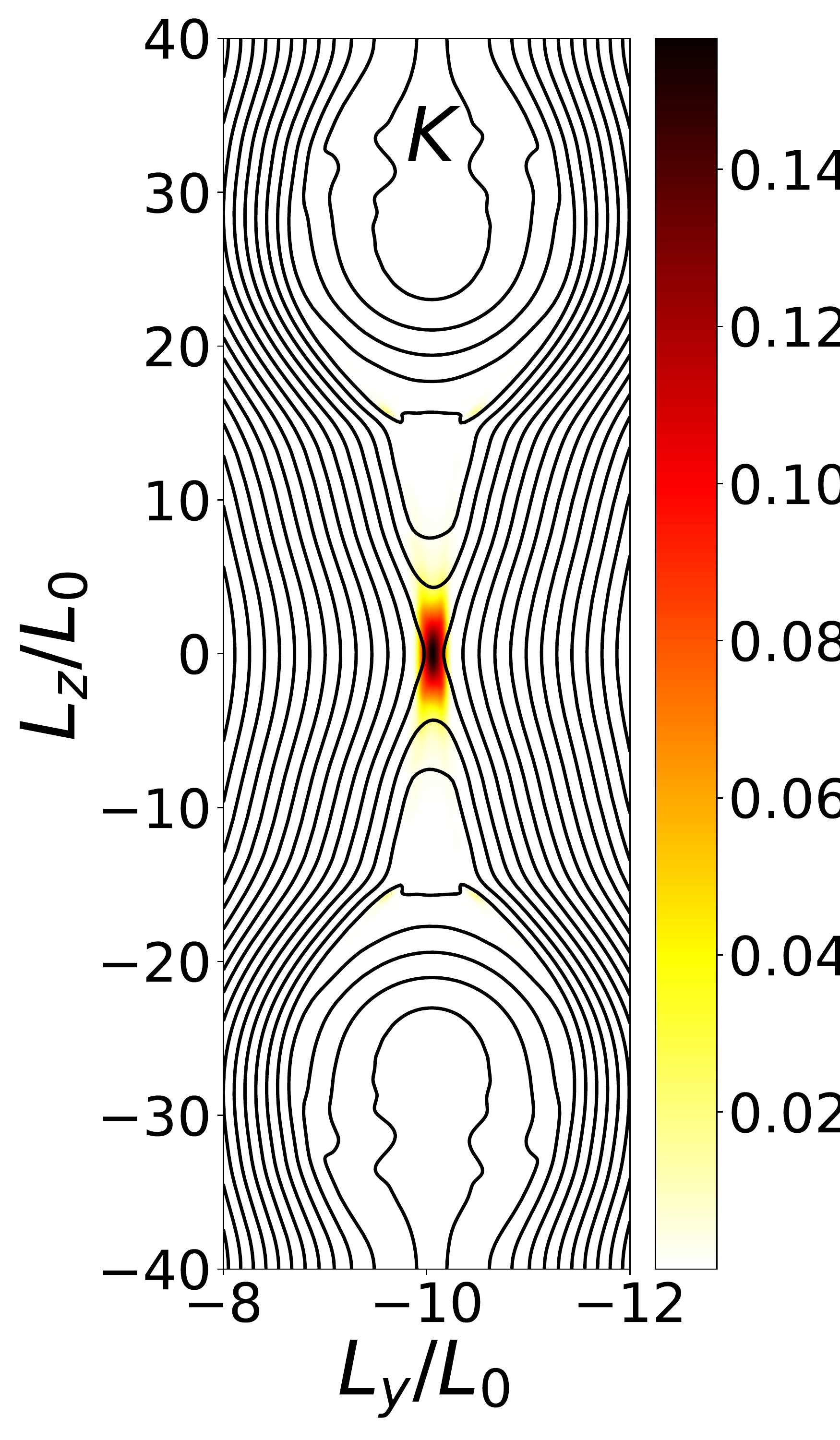} 
	\caption{Turbulent energy $K$}
	\label{fig:KEp}
       \end{subfigure}~
       \begin{subfigure}[t]{0.5\linewidth}
	\centering
	\includegraphics[width=0.9\textwidth,keepaspectratio]{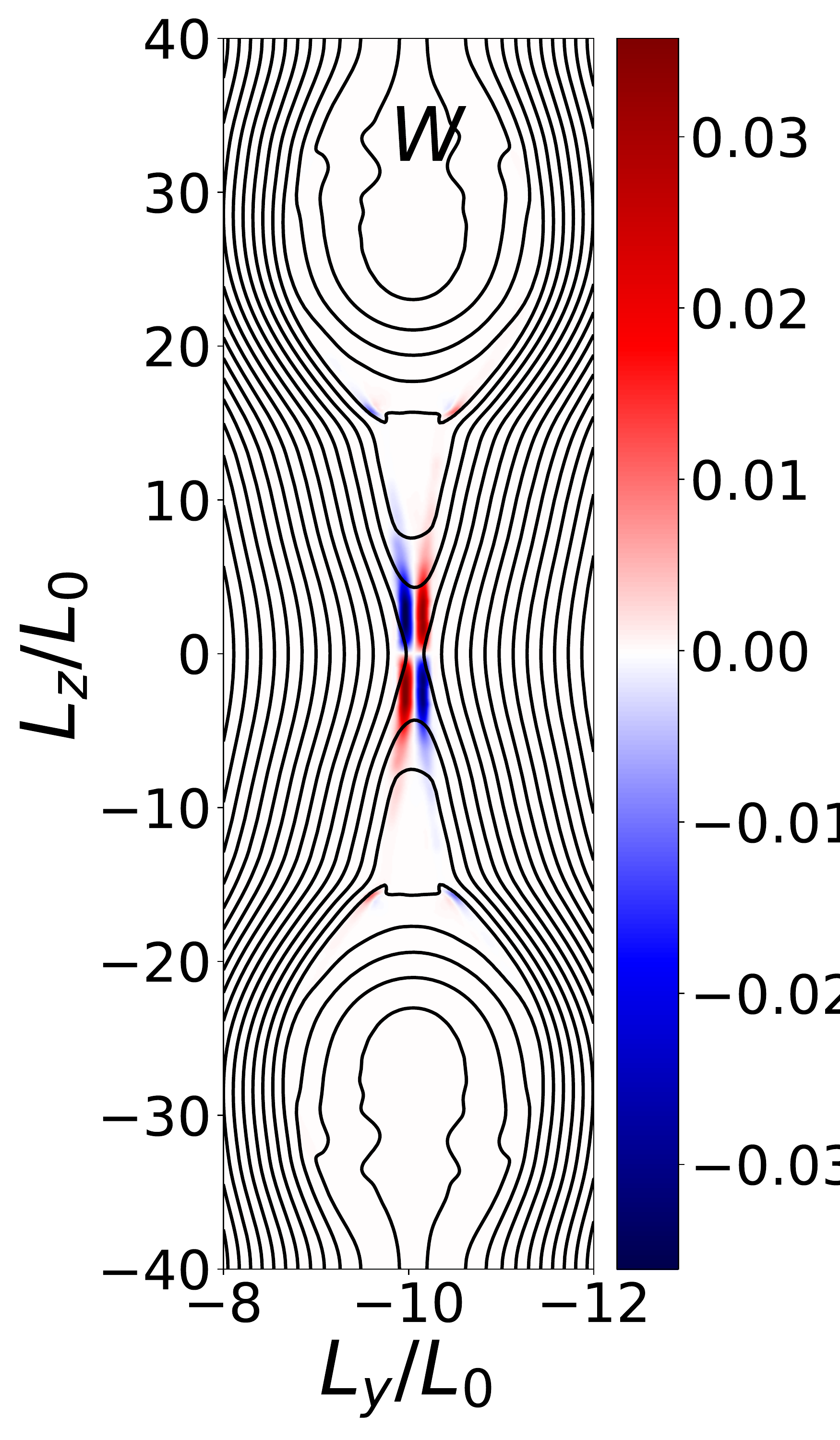}
	\caption{Turbulent cross-helicity $W$}
	\label{fig:WEp}
       \end{subfigure}\\
       \begin{subfigure}[t]{0.5\linewidth}
	\centering
	\includegraphics[width=0.9\textwidth,keepaspectratio]{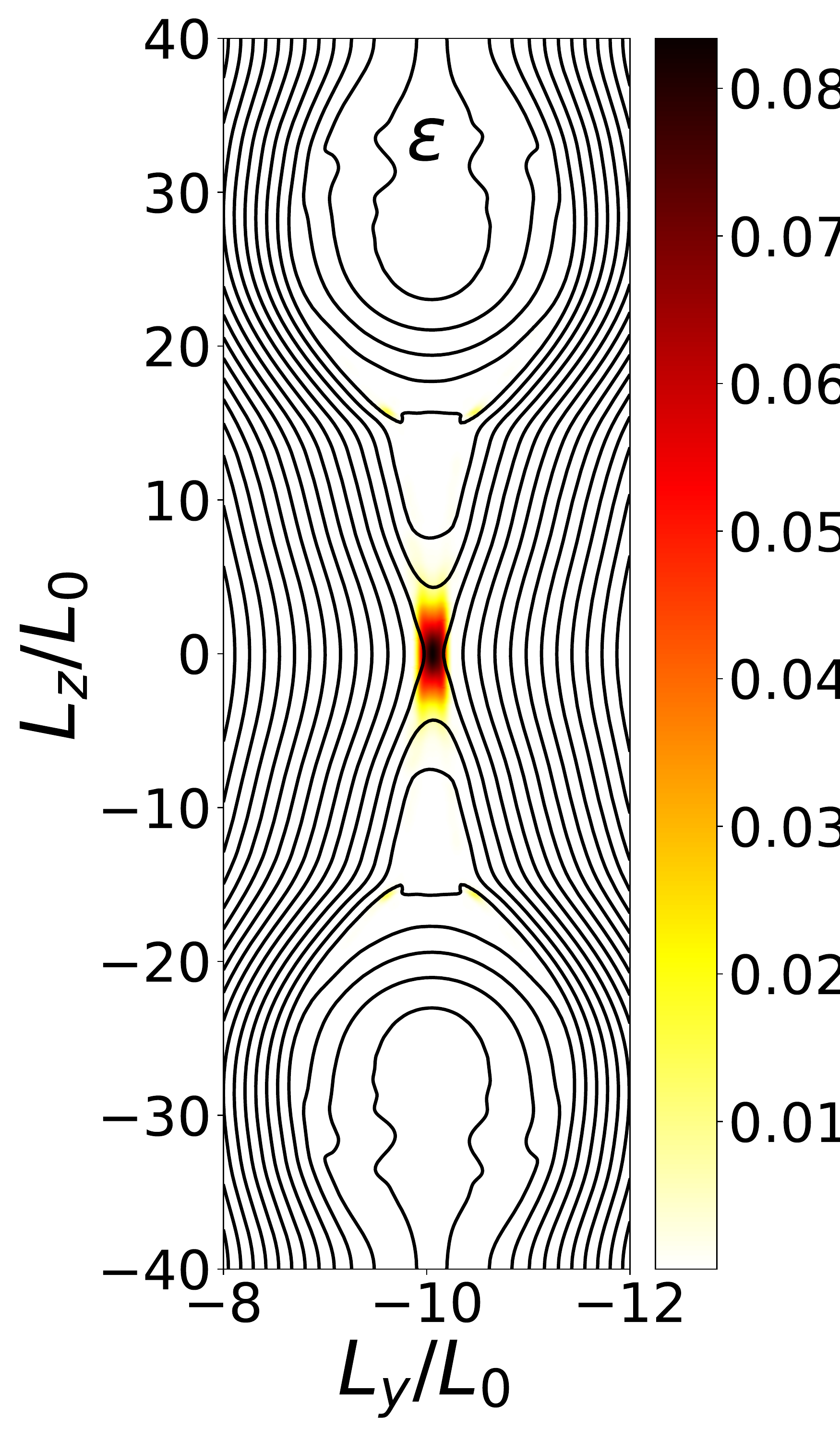}
	\caption{Dissipation rate $\varepsilon$}
	\label{fig:Epsi}
       \end{subfigure}~
       \begin{subfigure}[t]{0.5\linewidth}
	\centering
	\includegraphics[width=0.88\textwidth,keepaspectratio]{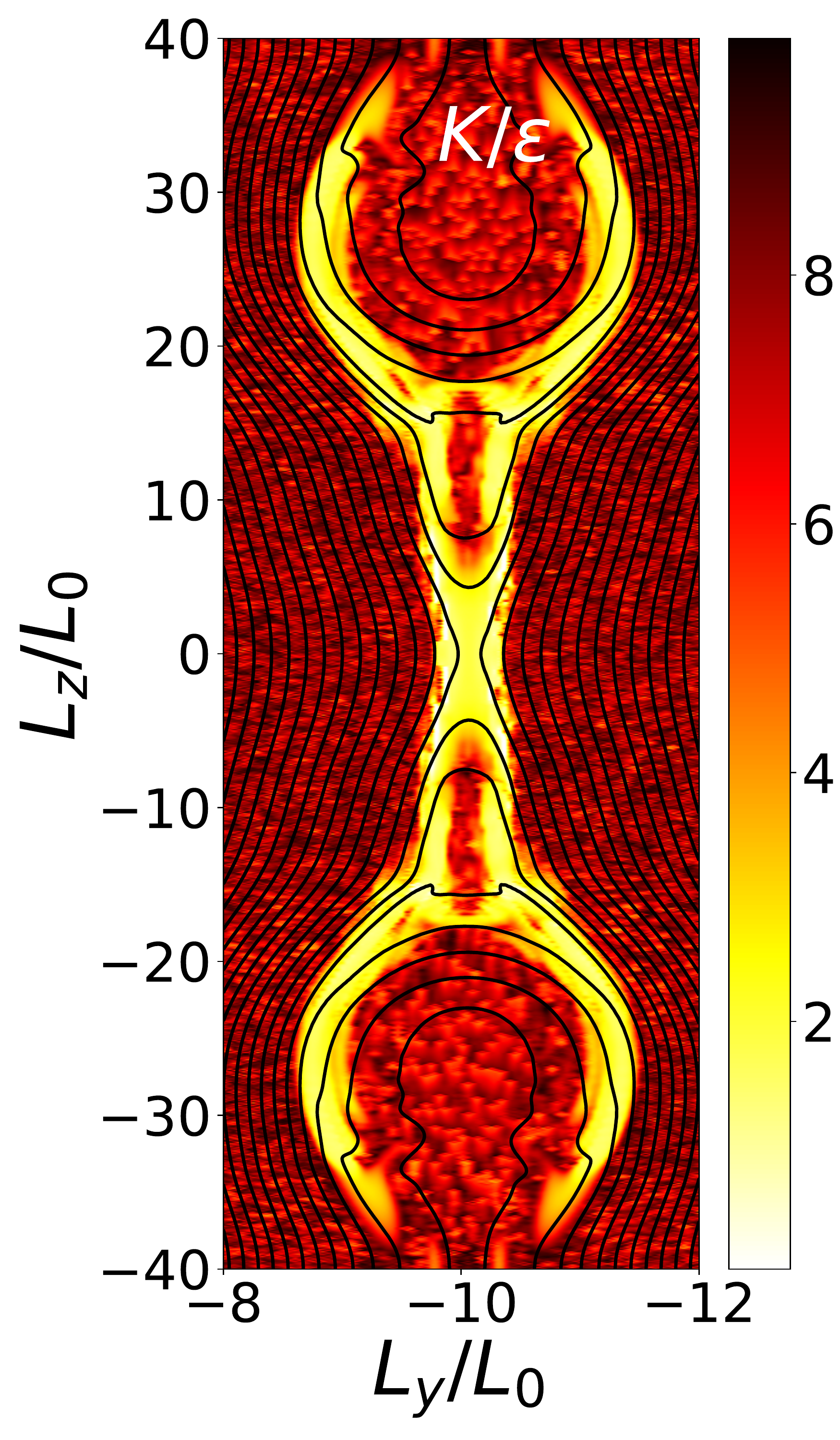}
	\caption{Turbulent timescale $\tau=K/\varepsilon$}
	\label{fig:Tau}
       \end{subfigure}
	\caption[Spatial distributions of  \ref{fig:MeanJEp} the mean current density, (b) the mean vorticity, (c) the turbulent energy, (d) its dissipation rate, (e) the turbulent cross-helicity, and (f) the turbulent timescale.]{Spatial distributions of (a) the mean current density $\boldsymbol{J}$,(b) the mean vorticity $\boldsymbol{\Omega}$, (c) the turbulent energy $K$, (d) the turbulent cross-helicity $W$, (e) turbulent energy dissipation rate $\varepsilon$, and (f) the turbulent timescale $\tau=K/\varepsilon$ at the time the reconnection rate saturates $(t=150\tau_A)$. $K_0=0.01$ The resistivity is $\eta=10^{-4}$.}
\label{fig:SpatialJKEW}
\end{figure}
\begin{figure}
	\centering
	\includegraphics[width=0.5\textwidth,keepaspectratio]{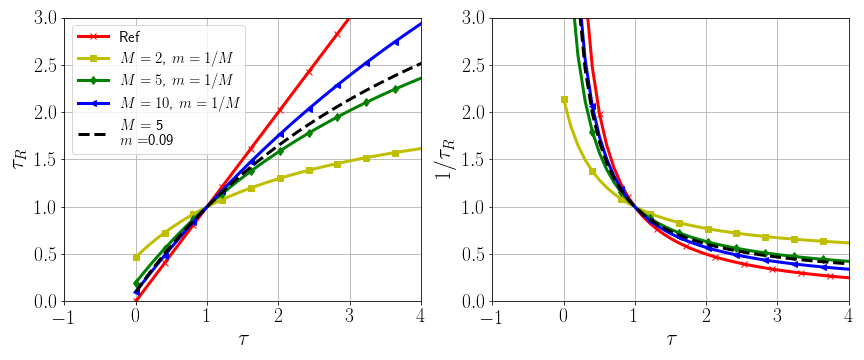} 
	\caption{Regulated turbulent timescale $\tau_R$ as a function of $\tau$ for various regulators.}
	\label{fig:RegulTau}
\end{figure}

The turbulent timescale $\tau$ is regularized to avoid numerical errors as\cite{MagetNum}
\begin{equation}
	\tau_R=\left(\frac{1}{M}+\frac{1}{1+m}\right)/\left(\frac{1}{M}+\frac{1}{m+\tau}\right),
	\label{eq:Regul}
\end{equation}
where $M$ and $m$ are the maximum and minimum values that the timescale of turbulence $\tau$ can attain. Since the turbulent timescale is normalised to the Alfvén time $\tau_A$ in our simulations, both limiters avoid unphysical values of $\tau$. \Fref{fig:RegulTau} represents \eref{eq:Regul} for
different values of $M$ and $m$. In previous work, the algebraic timescale was considered through the parameter $C_t=\tau_0/\tau$ where $\tau_0=C_\beta^{-1/2}|\boldsymbol{J}|_{t=0,z=0}^{-1}$. We choose $m=0.06$ and $M=10$ to let $\tau$ the
possibility to span reconnection regimes from laminar to turbulent as well as turbulent diffusion obtained within the range $C_t\in[0.05,3]$.\cite{Yokoi3,Widmer1}

\section{Fast Reconnection}
\subsection{Reconnection rate}
	The simplest theory of reconnection is the Sweet--Parker theory based on the conservation of magnetic flux and matter.\cite{JGR:JGR677} The model, developed for  resistive-MHD (hereafter $\eta$-MHD), assumes that two regions exist around the current sheet. Magnetic reconnection is then estimated at the time the current sheet has reached a steady state, i.e., when the diffusive term ($\eta$) of the induction equation balances the convective term ($\boldsymbol{V} \times \boldsymbol{B}$). The estimated reconnection rate ($M_{\rm{SP}}$) is found to be proportional to the inverse of the square root of the magnetic Reynolds number 
\begin{equation}
	M_{\rm{SP}}=Rm^{-1/2}\equiv\sqrt{\frac{\eta}{VL}}
	\label{eq:SP}
\end{equation}
with the resistivity $\eta$, the typical length scale $L$, and the plasma velocity $V$. This simple form is not a good approximation for large-magnetic-Reynolds-number astrophysical plasma. Using our mean-field turbulence model, the magnetic reconnection rate can be approximated by replacing $\eta\rightarrow\eta+\beta$ providing a turbulent magnetic Reynolds number $(Rm^{\rm{T}})^{-1/2}\equiv\sqrt{(\eta+\beta)/(VL)}$ for the Sweet--Parker model. Such a simple estimation shows that for collisionless astrophysical (large-magnetic-Reynolds-numbers) plasmas, turbulence might enhance the magnetic reconnection rate by a factor $\sqrt{1+\beta/\eta}$ above the $\eta$-MHD estimation.

	\Fref{fig:SpatialJKEW} depicts the spatial distributions of the mean current density $\boldsymbol{J}$, the mean vorticity $\boldsymbol{\Omega}$, the turbulence cross-helicity $W$, the turbulence energy $K$, its dissipation rate $\varepsilon$, and the turbulence timescale $\tau$ estimated from the evolution of $K$ and $\varepsilon$. We found that the turbulence energy $K(\propto\beta)$ is acting as a localized anomalous resistivity located at and around the `X'-point. Also, the turbulence energy dissipation rate $\varepsilon$ is finite at and around the diffusion region where the turbulence energy $K$ is maximum. The location where $\varepsilon$ is finite also represents the region in which the large scale magnetic field energy is transported to smaller scales where it can be more effectively dissipated. The timescale of turbulence $\tau$ is therefore maximum near the diffusion region where magnetic reconnection takes place. The intensity of $\tau$ is in the range $[1.2;1.4]$ corresponding to the regimes of \textit{fast turbulent reconnection} obtained for a constant turbulence timescale.\cite{Yokoi3,Widmer1}

	In previous works, a regime of slow energy conversion similar to the $\eta$-MHD rate (\textit{laminar reconnection}) and that of turbulent diffusion with slower rate of energy conversion (\textit{turbulent diffusion}) were obtained as well as the regime of fast reconnection (\textit{turbulent reconnection}).\cite{Yokoi3,Widmer1} These regimes were consequences of variation of an adjustable timescale parameter $\tau$ or the initial turbulence energy intensity $K_0$ as being independent parameters. In this work, only the initial intensity of the turbulence energy $K_{\rm{init}}$ is varied because the counterpart of the energy dissipation rate, $\varepsilon_0$, is directly related to $K_0$ by \eref{eq:Epsi0}. The model constants $C_\beta$, $C_\gamma$, $C_n (n=1-3)$ obtained from the turbulence modeling are not varied. Also, the current-density intensity is determined by the Harris-type current sheet initialization. 

\begin{figure}[h!]
  \centering
  \includegraphics[width=0.75\linewidth,keepaspectratio]{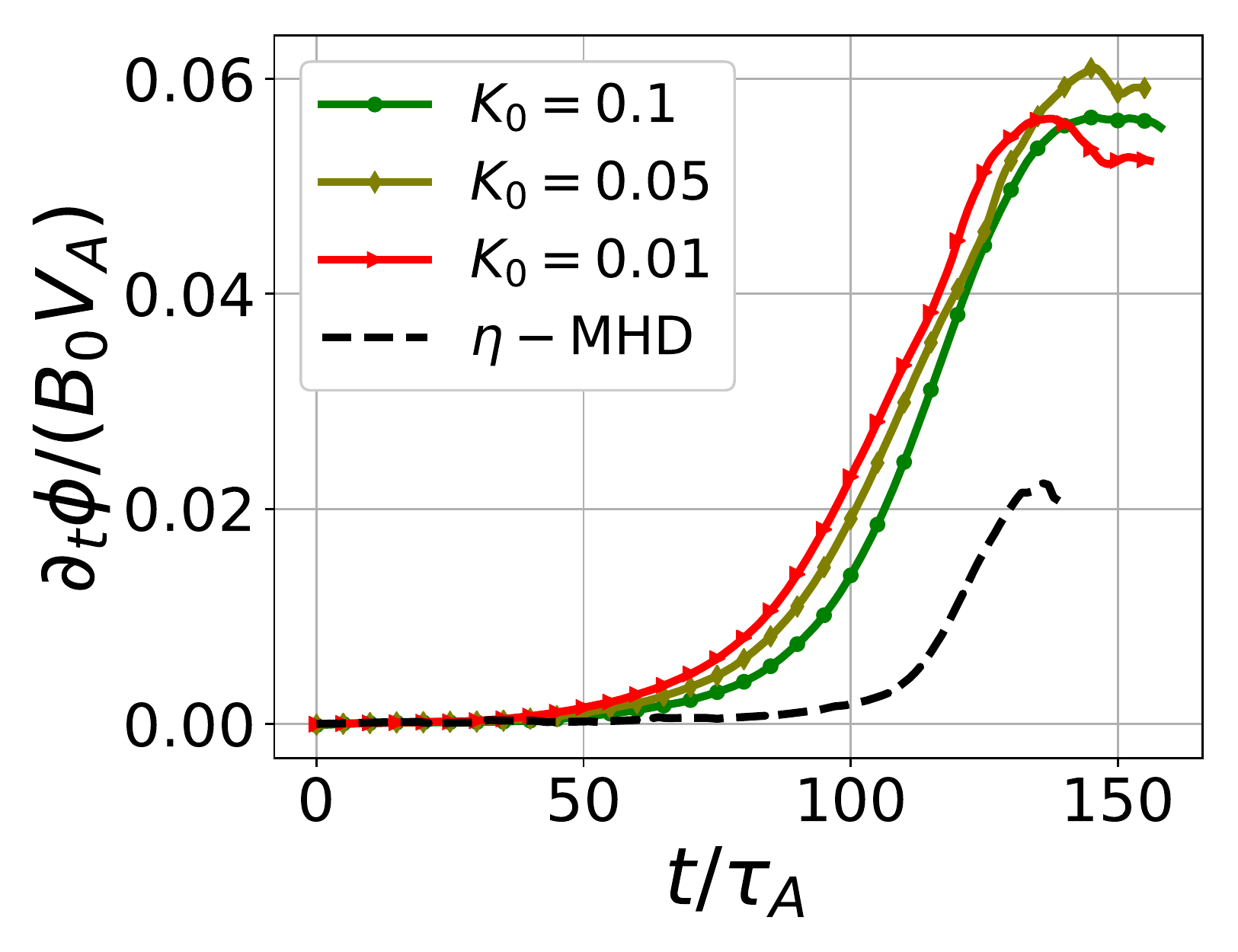}
  \caption{Time history of the reconnection rate for different initial values of the turbulent energy $K_0$ for $\eta=10^{-4}$.}
  \label{fig:RecRateEspiK}
\end{figure}

	\Fref{fig:RecRateEspiK} presents the time evolution of the reconnection rate as $K_{\rm{init}}$ is varied for a resistivity $\eta=10^{-4}$. Turbulence produces fast reconnection in comparison to the $\eta$-MHD regime.

Because the turbulence diffusivity ($\beta$) is localized, the Petschek model of reconnection describing a localized diffusion region due to standing shock waves\cite{1964NASSP..50..425P} is a more appropriate estimation of the magnetic reconnection rate. According to the Petschek model, the reconnection rate ($M_{\rm{P}}$) is expected to be proportional to the inverse of the logarithm of the magnetic Reynolds number as
\begin{equation}
	M_{\rm{P}}\approx\frac{1}{\log(Rm)} \sim \frac{1}{\log{1/\eta}}
	\label{eq:Petschek}
\end{equation} 
Utilizing \eref{eq:Petschek}, the turbulent reconnection rate ($M_{\rm{T}}$) may be estimated as
\begin{equation}
	M_{\rm{T}}\approx\frac{\log{\eta}}{\log{(\beta+\eta)}}M_{\rm{P}}\approx3.2 M_{\rm{P}},
	\label{eq:approxRR}
\end{equation}
with the intensity of $K$, $\varepsilon$, and $\tau=K/\varepsilon$ being taken at the reconnection peak (\fref{fig:SpatialJKEW}).

\begin{figure}
  \centering
  \includegraphics[width=0.75\linewidth,keepaspectratio]{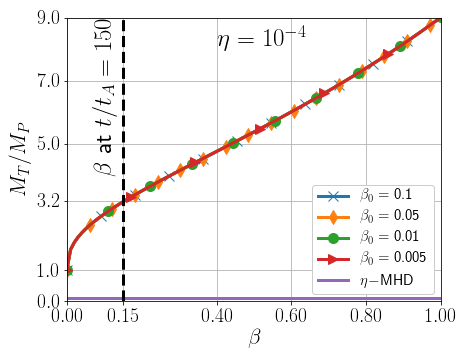}
	\caption{\eref{eq:approxRR} for various $K_0$.}
  \label{fig:MTMPTH}
\end{figure}
\begin{figure}
  \centering
  \includegraphics[width=0.75\linewidth,keepaspectratio]{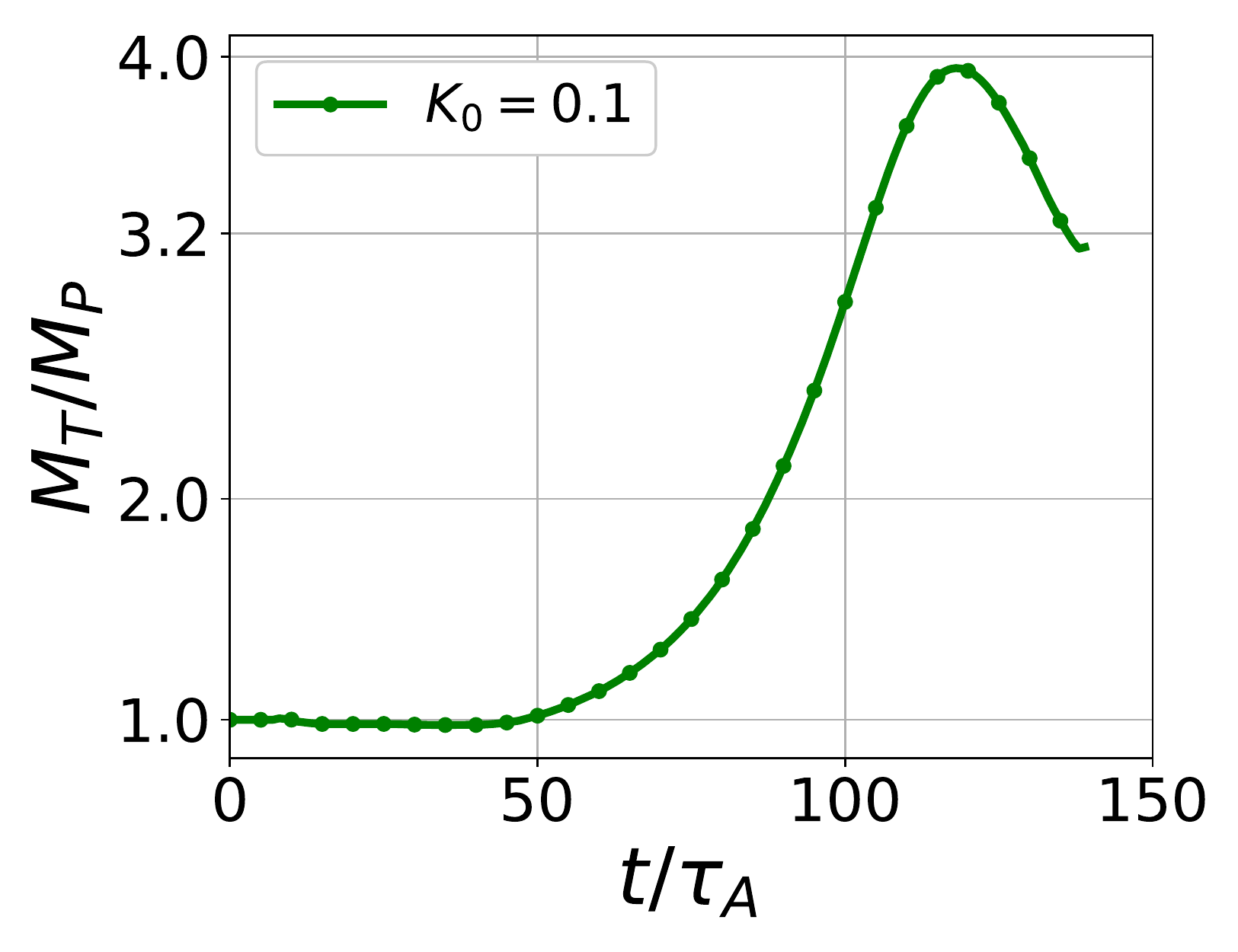}
	\caption{Time ratio of turbulent to $\eta-$MHD reconnection rate $K_0=0.1$.}
  \label{fig:MTMPT}
\end{figure}
\begin{figure}
  \centering
  \includegraphics[width=\linewidth,keepaspectratio]{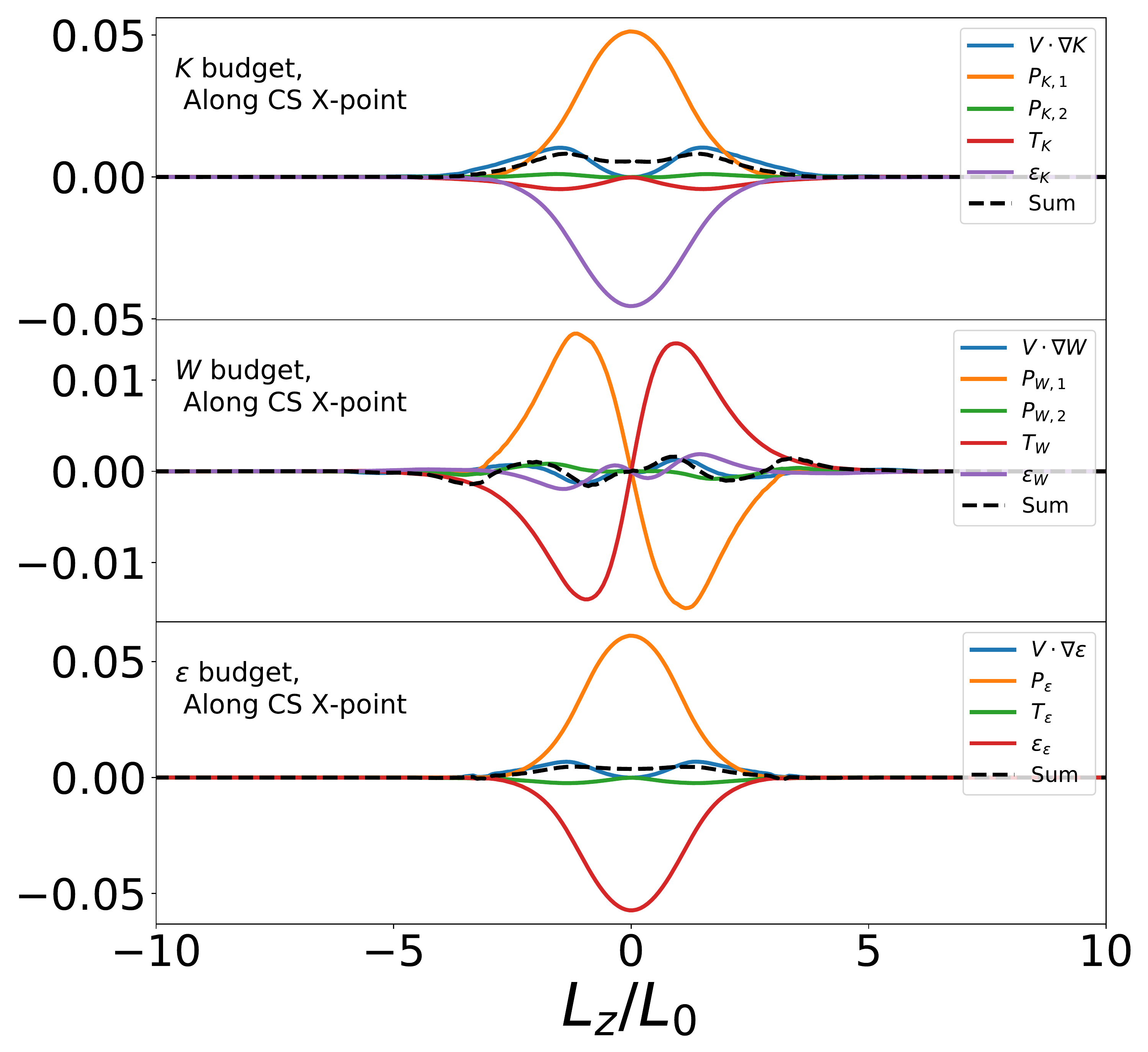}
	\caption{Contributing terms of \erefs{eq:KEpsi}{eq:Epsi} at $t/t_A=120$, $\eta=10^{-4}$.}
  \label{fig:KWEterms}
\end{figure}

\Fref{fig:MTMPTH} represents \eref{eq:approxRR} for various $K_0$. A turbulent energy intensity of $K=0.15$ enhances the $\eta-$MHD reconnection rate by a factor of about $3.2$ independently of the initial turbulence intensity $K_0$. \Fref{fig:MTMPT} shows time history of the turbulent reconnection rate to the $\eta-$MHD ratio. The ratio $M_T/M_P$ is about $3.2$ at the time both regime reach their saturation state. Turbulence enhances the
reconnection rate for large magnetic Reynolds number above the $\eta-$MHD value as a Petschek-like reconnection. 
In marked contrast to the previous results obtained from the simulations with a constant timescale parameter $\tau$, only \textit{fast turbulent reconnection} is obtained. In fact, the other two regimes of energy conversion, \textit{laminar reconnection} and \textit{turbulent diffusion}, are not reproduced in the present work, where the timescale of turbulence is self-consistently solved within the mean-field MHD equations. In this sense, the other two regimes, \textit{laminar reconnection} and \textit{turbulent diffusion}, are artifacts arising from the turbulence timescale as a constant parameter. 
\Fref{fig:KWEterms} depicts the production $P_{A,i}$, transport $T_{A}$, advective $\boldsymbol{V}\cdot\nabla A$ and dissipation $\epsilon_A$ terms in \erefs{eq:KEpsi}{eq:Epsi} at $t/t_A=120$. The reconnection rate initiates its saturation regime at this time. Here $A$ refers to $K$, $W$, or $\varepsilon$ and $i=1,2$.  The turbulent energy production term $P_{K,1}=C_\beta\tau K\boldsymbol{J}^2$ is located at the current sheet center and is mostly balanced by its dissipation term $\varepsilon_K$, a similar behavior is obtained for the turbulent energy dissipation rate $\varepsilon$. The evolution of turbulence reaches a steady state resulting in a saturation of the reconnected magnetic flux amount.
\section{Remark on Strong Compressibility Limit}
\begin{figure}
  \centering
  \includegraphics[width=0.85\linewidth,keepaspectratio]{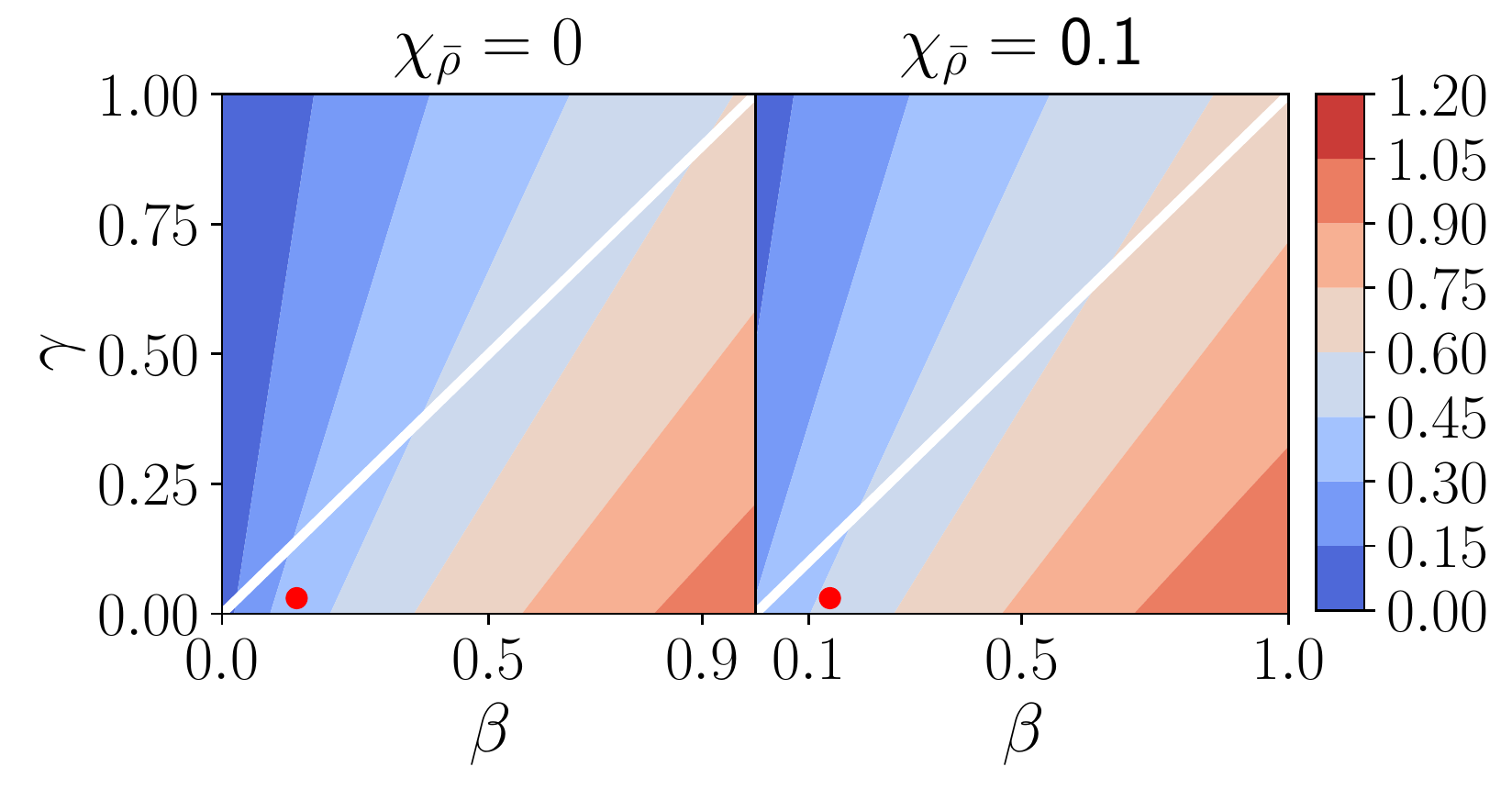}
	\caption{Sweet--Parker-like reconnection rate \eref{eq:SP_TH} as a function of $\beta$, $\gamma$ and $\chi_{\bar\rho}$. The red dots correspond to the maximum values of $\beta$ and $\gamma$ at $t=120\tau_A$.}
  \label{fig:SP_TH}
\end{figure}

The strong localisation of the turbulent energy diffusive-like term $\beta$
around the reconnection region produces a Petschek-like reconnection process.\cite{1964NASSP..50..425P}
In such a situation, slow shock-waves contributing to the fast Petschek-like
reconnection are usually associated with large density variance at the
reconnection region. We thus consider a contribution of the density variance in the electromotive force \mbox{\boldmath$\cal{E}$} (\eref{eq:Emf})\cite{Yok_Dens1,Yok_Dens2}
\begin{equation}
	\langle{{\boldsymbol{v}}'\times{\boldsymbol{b}'}}\rangle 
	= \alpha \boldsymbol{B}
	- \beta \mu \boldsymbol{J} 
	+\gamma {\bf{\Omega}}
	-\chi_{\bar\rho}\nabla\bar\rho\times\boldsymbol{B},
	\label{eq:Emf_Density}
\end{equation}
where $\chi_{\bar\rho}$ is the transport coefficient associated with the
density variance $\langle{\rho'{}^2}\rangle$ representing genuine compressibility effects. It is
important to stress that the last term in \eref{eq:Emf_Density} has no
counterparts in the incompressible case.
The derivation of the electromotive force in a strong compressible regime as
well as the physical origins of the $\chi_{\bar\rho}$ related term can be
found in Yokoi 2018.\cite{Yok_Dens1,Yok_Dens2} There, it is shown that the $\bf{B}\times\nabla\bar\rho$ direction contributes to the turbulent electromotive force, enhancing the
turbulent energy production across the reconnection slow
shock front.
Such a phenomena might increase magnetic reconnection. A simple consideration
of a Sweet--Parker current sheet  for the electromotive force \eref{eq:Emf_Density}, neglecting
the $\alpha$ related term, lead to the following reconnection rate
\begin{equation}
	M_{T,SP}^2=\frac{1}{2}\sqrt{4\left(\hat\eta+\hat\beta +\hat\chi_{\bar\rho}\right)+\hat\gamma^2}-\frac{\hat\gamma}{2},
	\label{eq:SP_TH}
\end{equation}
the hats denoting normalised variables .
\Eref{eq:SP_TH} is represented in \fref{fig:SP_TH} for $\hat\chi_{\bar\rho}=0$, $0.25$. The red dot represents the values of $\beta$ and $\gamma$ at $t=120\tau_A$
and the white line represents the limit of turbulence saturation for fast
reconnection $|\gamma|/\beta=1$.\cite{Yokoi2} We see that this simple estimate
underestimate the reconnection rate obtained through our full-compressible MHD
numerical simulations for $\chi_{\bar\rho}=0$ but is in a similar range when $\chi_{\bar\rho}=0.1$. Including the density variance effect in the turbulent electromotive
force \mbox{\boldmath$\cal{E}$} enhances the reconnection rate.
Further numerical validations based either on Direct Numerical Simulations (DNSs) or turbulent mean-fields simulations evolving evolution equations for turbulent transport coefficients, as done previously,\cite{Widmer2,Yokoi3,Widmer1} are required.

\section{Conclusions}


	In this work, we have performed numerical simulations of a turbulence model for magnetic reconnection with an equation of the turbulent energy dissipation rate ($\varepsilon$ equation) implemented.\cite{Yokoi1} In the simulations, where the timescale of turbulence is evaluated from the turbulent energy $K$ and its dissipation rate $\varepsilon$, we obtained only the regime of \textit{fast reconnection} among the three regimes of the turbulent reconnection.\cite{} From the viewpoint of the mean-field turbulence model, this is quite a natural result. The turbulence timescale is determined by the nonlinear dynamics through the coupling of the evolutions of the turbulence energy $K$ and its dissipation rate $\varepsilon$. The contribution of the production and
dissipation terms in the evolution equations of $K$ and $\varepsilon$ are competing. They mostly
balance close to the time of reconnection rate saturation, turbulence production and dissipation reach a steady state and cannot enhance the reconnection process further.\newline\indent
We found that the magnitude of the turbulence energy $K$ is of the same order as the counterpart of the turbulent dissipation rate $\varepsilon$, and that the timescale provided by the ratio $\tau=K/\varepsilon$ is about twice unity in the diffusion region. The value of $\tau$ corresponds to the time needed for an Alfvén wave to cross the initial current sheet width ($\tau_{\rm{A}} \approx \tau_{\rm{AE}}$). In such situations, the eddy-turnover time can be used to estimate the timescale of turbulence as $\tau_{\rm{E}} = K / \varepsilon$.
As this result, the rate of energy conversion of our mean-field turbulence model is enhanced above the rate of energy conversion of $\eta$-MHD.

	

\vspace{15pt}
\begin{acknowledgments}
	One of the authors (FW) acknowledges the International Max-Planck Research School (IMPRS) at the University of G\"ottingen as well as the CRC 963 project A15. JB thanks the Max-Planck--Princeton Center for Plasma Physics for its support. A part of this work was performed during the period when one of the authors (NY) visited Goslar and G\"{o}ttingen in August 2017, and under the support of the JSPS Grants-in-Aid for Scientific Research 18H01212. 
\end{acknowledgments}


\bibliography{References}

\end{document}